# Monte Carlo Simulations Demonstrating Physics of Equivalency of Gamma, Electron-beam, and X-ray for Radiation Sterilization

Thomas K. Kroc, PhD
Fermi National Accelerator Laboratory
PO Box 500, Batavia, IL 60510, USA
kroc@fnal.gov

## 1. Abstract

The sterilization of medical devices using the gamma rays from the decay of cobalt-60 has accumulated decades of experience of the performance of the materials and devices that are irradiated. The use of radiation using electron beams and x-rays has much less experience and this leads to questions of equivalency between these three technologies. Computer simulations were conducted to model the relevant physical processes of the interactions of each of the three forms of radiation in order to compare the spectra of electron energies at energies below 500 keV. It is predominantly the electrons below this threshold that produce the sterilization dose. No difference in energy spectra was seen between the three types of initial radiation. It is concluded that there is no energy dependent difference between gamma, electron, and x-ray for radiation sterilization.

## 2. Introduction

An important issue in radiation sterilization concerns the interchangeability of the three radiation sources: gamma rays from the decay of Cobalt-60, x-rays produced from an electron beam striking a bremsstrahlung converter, and direct electron beams. While gamma rays from the decay of cobalt-60 have been used for medical device sterilization for decades, there is much less experience with electron beams and x-rays. This has led to uncertainty as to the equivalence of electron beams and x-rays with gamma rays. The work reported here seeks to demonstrate the equivalence of the active element in radiation sterilization, that is low energy electrons, from all three radiation sources. It is important to note that ISO 11137 [1] acknowledges and accepts all three forms of radiation production. Similarly, guidance documents such as the newly released AAMI TIR 104 [2], accept all three radiation sources.

Large scale irradiation facilities treat product in bulk. The product can be transported through the irradiation chamber on pallets, by the use of carriers called totes, or, mainly in electron facilities, in individual shipping boxes or cartons. For the purposes of estimating the irradiation time to accrue the desired dose, an average density of the irradiated container is assumed. The delivered dose is verified by placing dosimetry at various points on and inside the product containers. Electrons of energies of tens to thousands of electron volts (eV), but much less than the initial particle energies of millions of eV, are responsible for depositing dose. Computer simulations have been conducted for this report and have been set up to simulate the geometry of actual irradiation enclosures. An average density of the irradiated product has been used. The object is to observe whether or not there is a difference in the electrons that are generated in the interaction of the radiation from the three sources with the simulated product. If there are no differences, then it can be assumed that irradiated materials will behave similarly regardless of the radiation source. Consequently, the sterilizing power of each form of radiation should be the same.

There is a slowly growing body of literature that compares the three sources of radiation. Some have compared two or three of these sources of radiation [3,4]. Others have analyzed the safety of the higher energy sources; x-ray to energy up to 7.5 MeV and electron beams up to 10 MeV [5,6]. A consortium of national laboratories, industry, and academia is investigating the properties of materials used in medical devices using all three sources of radiation [7,8,9, 10]. Some literature examines the microbial effects of various sources [11]. Still other literature examines the radiation chemistry which looks at the number of molecules or free radicals that are generated by various ionizing radiations [12]. This work supplements those resources by providing a microscopic look at the processes of the interaction of radiation with the material of an irradiated product.

In looking at the initial energy spectra for each radiation technology, Figure 1, it is easy to infer that there are important differences and that comparing their ability to sterilize materials would be difficult. Here however, appearances are deceiving, as these spectra are of the electrons or photons immediately after their emission. (The cobalt case includes a slight amount of internal scattering, the x-ray case shows the immediate result of the bremsstrahlung production, and the electron beam case includes only the small momentum spread of the electron accelerator.) They do not include scattering and attenuation effects that occur within the source structure, let alone the materials they encounter between the source assemblies and the product to be irradiated. This work will show that fundamentally, when one gets down to the point of the mechanisms that cause sterilization, the three radiation sources are indistinguishable from each other.

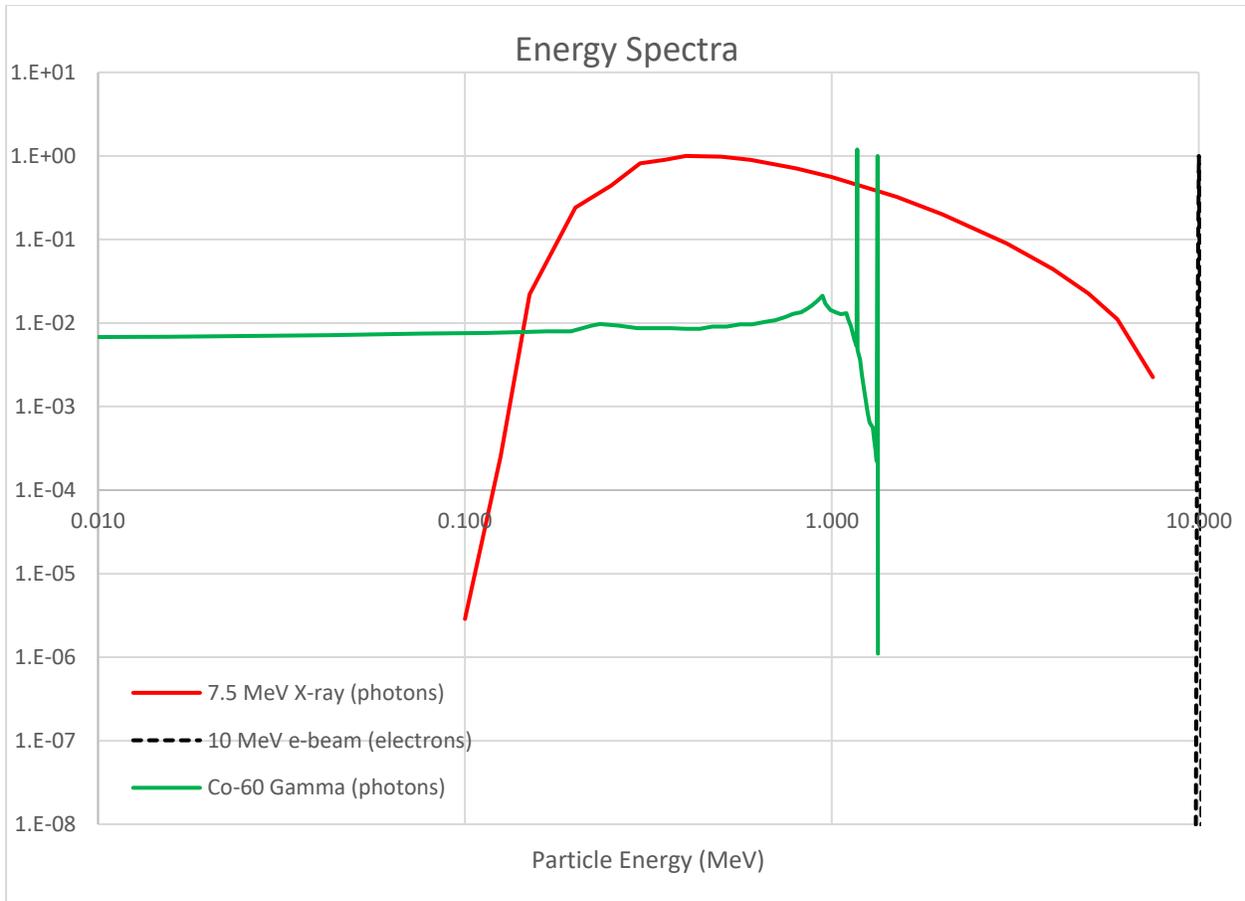

Figure 1 Energy spectra of photons or electrons exiting their ideal sources for three radiation sources. X-ray spectrum generated using parameters in Ref. 13.

The energy of the source particles, photons or electrons, that we are considering are between 1 and 10 MeV. The minimum energy required to create ionization is on the order of 100 eV which is 10,000 to 100,000 times less than the energy of the source particles. When the source particles strike matter, say a device to be sterilized, they set off a cascade of photons and electrons that increase in number of particles while the energy per particle decreases as the cascade progresses. As the average energy per particle decreases, the number of ionizations increases proportionally. This continues until the energy per particle decreases below the energy needed to ionize an atom. At this point the remaining energy is deposited in the matter as heat. But it is this blooming of ionization events, initiated by each initial particle that is the sterilizing power of radiation.

We will investigate the spectra of the low energy electrons to see if there are any differences between the three radiation sources that might result in a difference in their operation.

## 3. Discussion
### 3.1 Interaction of Radiation with Matter - Macroscopic Description

Electrons are charged particles. As charged particles travel, they have a specific maximum distance, referred to as their range, which is energy dependent. During this travel, the energy that an electron deposits varies with depth and increases to a sharp peak at the end of its travels, known as a Bragg Peak [14]. This deposition, the linear energy transfer (LET), is plotted as a function of the traveled path in Figure 2, (dark solid line) and can be denoted as dE/dx. However, since electrons are so light compared to the atoms and molecules they are interacting with, they undergo many large scattering events which cause a buildup of the energy deposition at a distance that is approximately half of its maximum range. This is shown in the green dashed line Figure 2. Since it is difficult to measure the dE/dx in a linear manner directly, the dE/dx in Figure 2 was evaluated by integrating the calculated stopping power [15] of the electrons (in MeV/cm) as a function of the decreasing energy with distance.

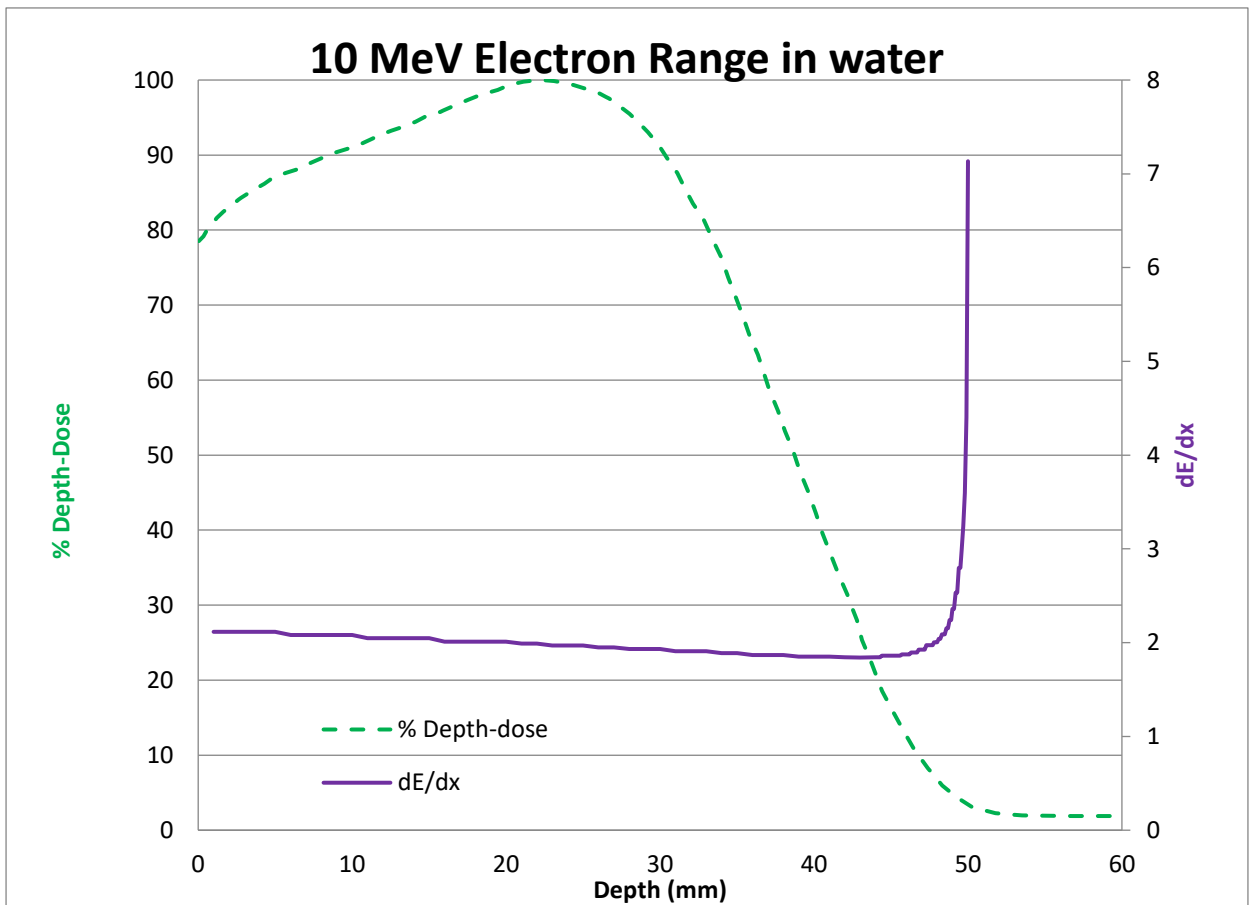

*Figure 2 A plot of the range of 9 MeV electrons in water and the linear energy transfer, dE/dx, of an electron if scattering of the trajectory of the electron did not occur (dark solid line). However, because of scattering, the terminal peak of each electron's energy deposition (the Bragg Peak) happens at a random depth which results in the maximum energy deposition peak of the conglomerate to occur at approximately 46% of its maximum range (dashed green line). (data for plot from [15].)*

Photons are uncharged particles. They do not have a definitive range. When a beam of photons is directed onto matter, the intensity of the beam is attenuated exponentially. This is shown in Figure 3 for photons from both cobalt-60 and 7.5 MeV bremsstrahlung x-rays, along with a 10 MeV electron beam for comparison.

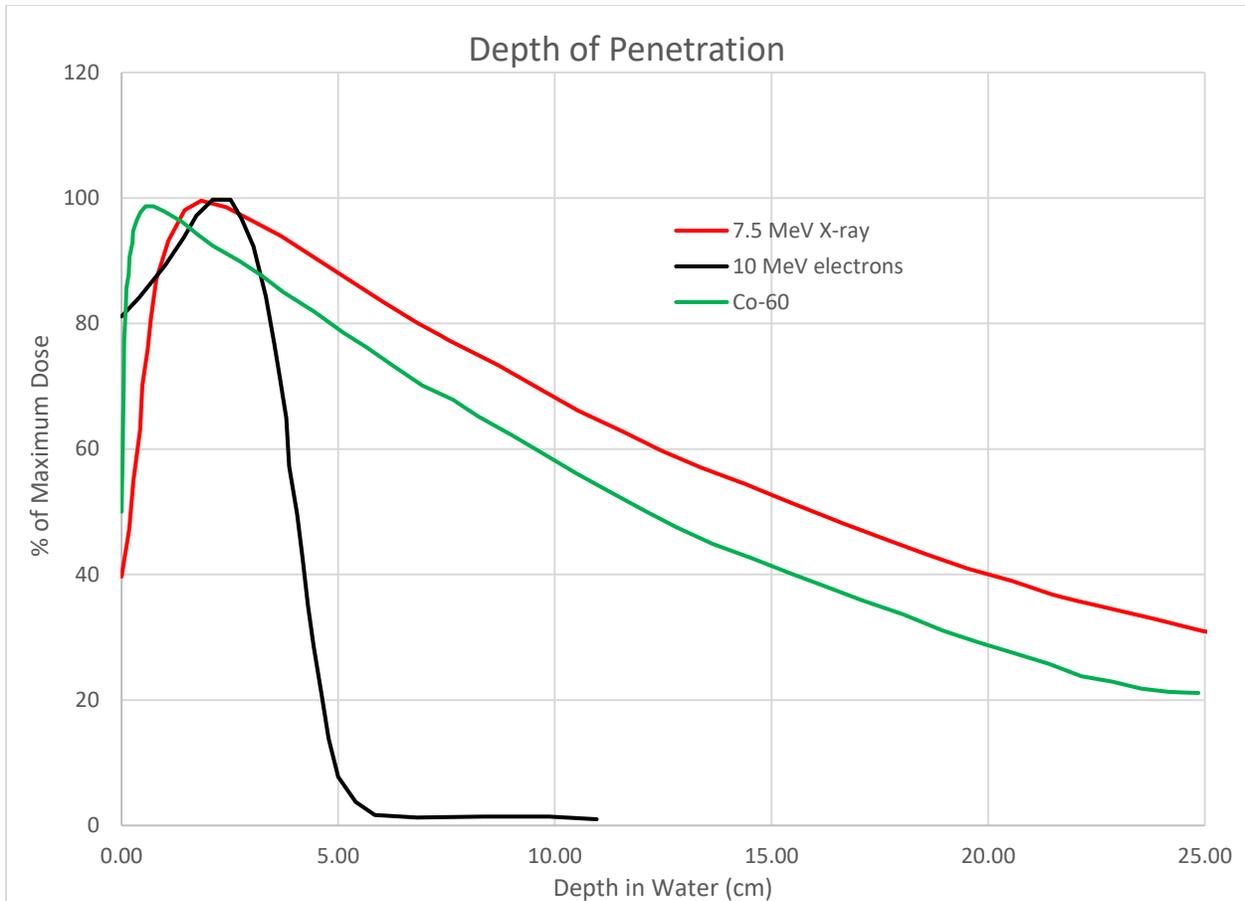

*Figure 3 The penetration of the three radiation sources as a function of depth in water.*

In industrial sterilization, it is normal to use the average density of a package to determine the dose penetration. In such a case, the distance scale would be extended inversely proportional by that density.

### 3.2 Interaction of Radiation with Matter - Microscopic Description

Dose is a measure of the energy absorbed in matter from ionization and is defined as the amount of energy, in joules (j), absorbed per unit mass in kilograms (kg). The unit of dose is the Gray (Gy). In industrial sterilization, the kiloGray is frequently used, kJ/kg = kGy. Charged particles (i.e., electrons), are defined as directly ionizing radiation as it is the interaction of the Coulomb forces between them and the electrons of the atoms of an irradiated material that creates ionization. Photons are defined as indirectly ionizing radiation as they must first generate a charged particle, an electron to initiate ionization. These processes build into a cascade, with a continual cycle of photons liberating electrons, electrons creating new photons, and electrons causing ionization until the energy has been dissipated.

   a. The Compton effect, the photoelectric effect, and pair production – photons beget electrons

When a high energy photon beam, and here we are thinking of energy of at least one MeV, encounters the matter that comprises the irradiated product, there are three primary interactions that can initially occur that release electrons: the Compton effect [16], the photoelectric effect [17], and pair production (the creation of an electron-positron pair) [18].

The Compton effect is actually a scattering process. The initial photon is deflected by an electron in an atom. During the deflection, some of the energy of the photon is transferred to the electron. If the amount of energy transferred is greater than the binding energy of the electron, it escapes from the atom and an ion is created. The diminished photon continues on a new trajectory and, assuming it still has enough energy, can create more ionizations. The average energy of an electron from the Compton effect is ~500 keV [19].

The photoelectric effect is a similar to the Compton Effect except that all the photon's energy is transferred to the electron and the incident photon ceases to exist.

If the energy of the photon is greater than or equal to 1.022 MeV, an electron-positron (anti-electron) pair can be produced in the vicinity of the nucleus of an atom. If the photon has energy greater than 1.022 MeV, then the excess energy is divided equally into the kinetic energy of the two particles.

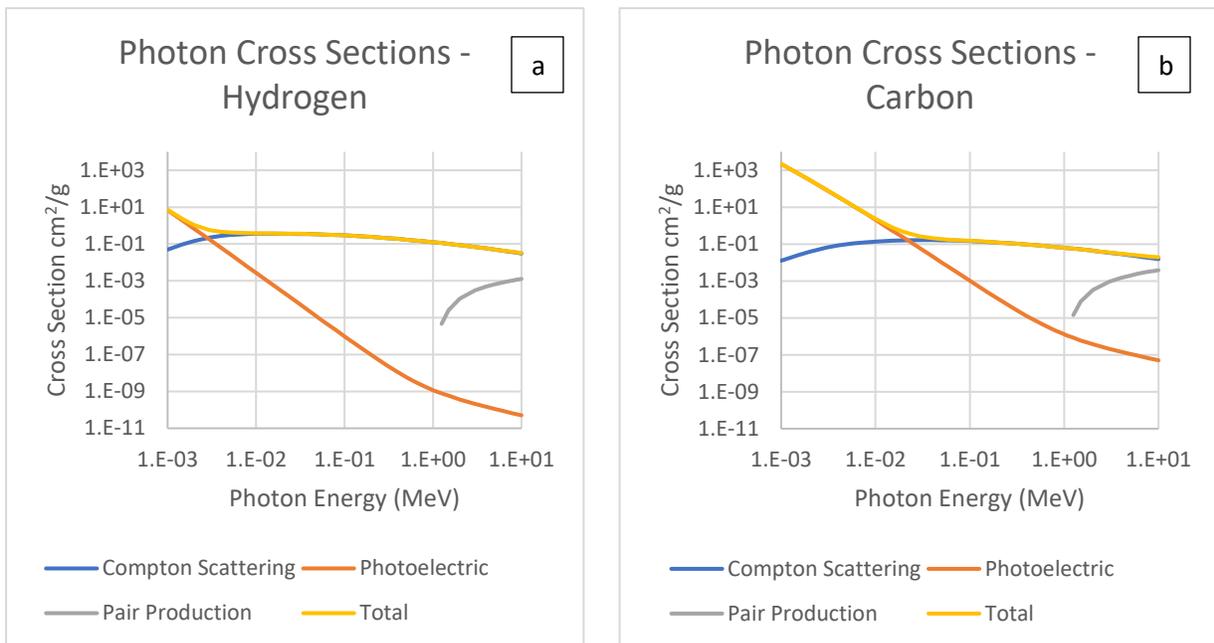

*Figure 4 Cross sections of Hydrogen (a) and Carbon (b) to absorb a photon due to three processes: Photoelectric Effect, Compton (Incoherent) Scatter, and Pair Production. For hydrogen, Compton Scattering predominates from about 2 keV to over 10 MeV; for carbon, Compton Scattering starts at about 20 keV. Other elements exhibit similar behavior although the transition between processes occurs at different energies. (Data from [20]) Note: the units of $cm^2/g$ are a method of normalizing these plots to unit density to better compare between materials.*

Figure 4 shows the mix of these three processes as a function of the photon energy on hydrogen (a) and carbon (b). Hydrogen and carbon are representative of the behavior of the light elements. They were chosen for this illustration since, in sterilization, we are interested in killing living organisms which are organic matter. In the range between 5 keV for hydrogen, 50 keV for carbon and 10 MeV for both, the Compton scattering is the predominant interaction mechanism.

### b. Bremsstrahlung – electrons beget photons

To produce an x-ray beam, a beam of electrons is directed onto a material with a high atomic weight, such as tantalum. When high-energy electrons encounter the electric field of a nucleus of an atom in this material, they experience an acceleration, a change in the direction of their trajectory. This results in a reduction of kinetic energy of the electrons and can cause photons to be emitted. Because the amount of deflection depends on the distance of closest approach and the initial energy of the electrons, the energy of the emitted photons do not have a specific energy and forms a continuous spectrum of energies with a maximum energy equal to the energy of the electron beam as was seen in the red curve of Figure 1.

### c. Lower energy processes

The bremsstrahlung process also occurs inside the material being irradiated. Medium to high energy electrons can create photons that then can interact through one of the three processes noted in Section 3.2.a. Other processes come into play as the cascade of events progresses. They are described here in rough order of decreasing energy. These are the processes noted in Figure 8 and Table 2.

Auger electrons can be created by either photons or electrons. Whereas in Compton scattering, the photoelectric effect, and pair production mainly involve outer shell electrons, interactions with inner shell electrons can also liberate electrons. Here, a vacancy is created in an inner shell. When an outer shell electron drops down to fill that vacancy, the difference in binding energy between the two levels can cause another electron to be ejected which is referred to as an Auger electron. Most Auger electrons travel only a few nanometers but have high values of dE/dx and therefore can be very potent against biologic material. [21]

Positrons, created in pair production interact with electrons in the material and annihilate each other resulting in a pair of photons each with energy of 0.511 MeV.

Whereas the bremsstrahlung process is an interaction with the nuclei of the irradiated material, the incident electrons can also create x-rays through interactions with the electrons of the irradiated material.

Fluorescence is a photon scattering phenomenon. Here, a photon is absorbed by an atom and reemitted at a lower energy.

At the terminal end of the cascade of events are knock-on or delta ray electrons. This is an electron-electron process where an incident electron scatters off another electron and liberates the second, causing an ionization.

While the magnitude of the energy of these individual events is lower than the energy of the initial radiation sources, the number of events has increased greatly and the distance between successive events is much smaller. Therefore, the density of the energy deposited by them is large. Therefore, these are the processes that have the most impact on biological material.

## 3.3 Putting It Together

Defining each of the processes that are involved does not give a complete picture of how the interaction of radiation and matter occurs. When a particle, either an electron or a photon strikes a substance, it triggers a cascade of interwoven events. Photons create or liberate electrons. Electrons produce photons. This goes back and forth with a continual reduction in the energy of the individual particles until the energy of those particles drops below the threshold for those processes. Each time an electron is liberated energy is being transferred to the impacted material which is the foundation of dose. Higher energy particles, particularly photons, but also to an extent, high energy electrons, serve to penetrate the object to be irradiated. As the various processes described above begin and the energy of the individual particles diminishes, the ionization processes described in Section 3.2.c take over.

Different disciplines focus on different aspects of these interactions. Atomic Physics (and the MCNP®(22) Monte Carlo program used for this work) track the processes noted above. Radiation Biology looks at the linear energy transfer . This compares the density of the energy transfer with the distance scale of biological structures to determine the biological effectiveness of different forms of radiation. Radiation Chemistry focuses on the creation of radical ions as the mechanism of action. This tends to look at the groupings of the energy deposited as noted below. To provide a visualization of the development of the interaction of radiation with matter, the radiation chemistry viewpoint will be used.

Figure 5 illustrates the variety of interactions that can occur. The left-hand side focuses on the high energy interactions described in Sections 3.1, 3.2.a, and 3.2.b above. The right-hand side illustrates the low energy interactions of the electrons described in section 3.2.c that occur at the terminal end of the electrons' paths. Radiation chemists view these interactions in terms of the energy deposited without identifying the actual interaction processes. The interactions are grouped by the amount of energy deposited as shown in the right-hand side of Figure 5. [23, 24] Spurs are defined to have transferred 100 eV or less. Blobs transfer between 100 and 500 eV. Short tracks transfer between 500 and 5,000 eV. These thresholds are associated with the mobility of the electrons relative to the ions left behind and the local density of ionizations. These interactions produce the bulk of the ionization and therefore are delivering dose. One can see that the energy of the interactions ranges between 100 and 5,000 eV, much below the energy of the source particles.

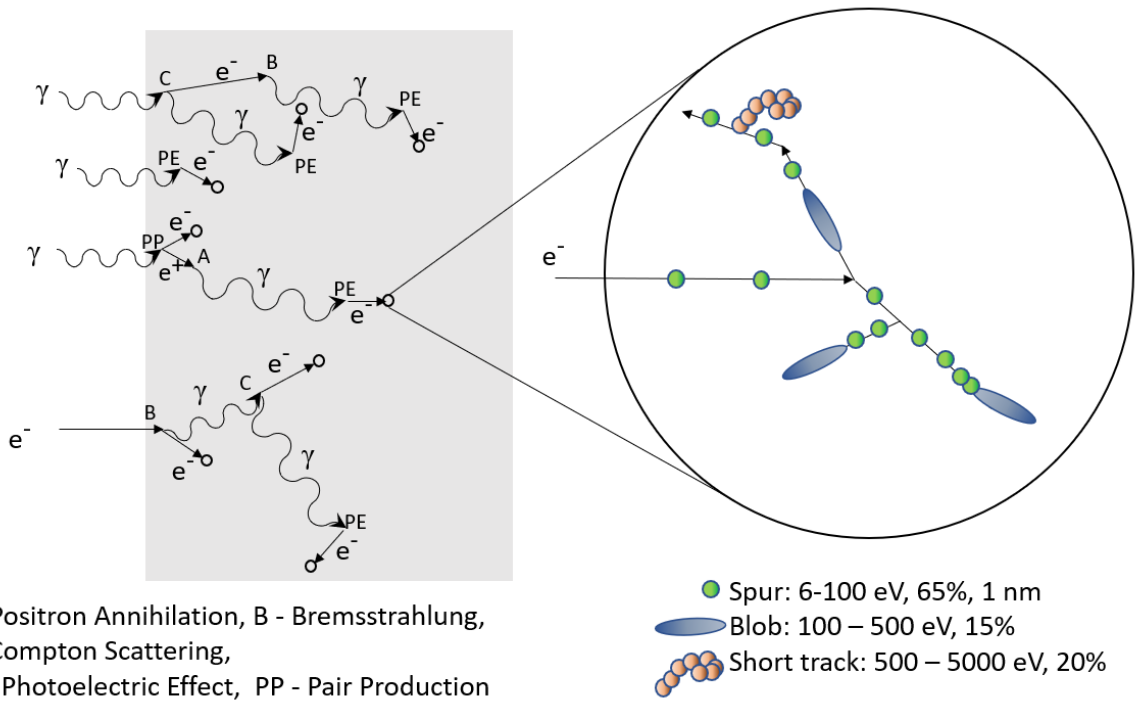

A - Positron Annihilation, B - Bremsstrahlung,
C - Compton Scattering,
PE - Photoelectric Effect,  PP - Pair Production

○ Spur: 6-100 eV, 65%, 1 nm
⬬ Blob: 100 – 500 eV, 15%
⚬⚬ Short track: 500 – 5000 eV, 20%

*Figure 5 Illustration of the interactions of electrons in matter creating ionization with notation of the typical energies of the interactions. All interactions end with electrons. Figure on right illustrates action at the terminal end of an electron's path.*

Switching to the Radiation Biology viewpoint and looking back at Figure 2, we can talk about the Relative Biological Effectiveness (RBE) or linear energy transfer (dE/dx) of the electrons. The RBE is an empirical value that is used to compare the biological effect of various forms of radiation. It is related to the linear energy transfer, which is the density of energy deposited, dE, along the path of a particle, dx. As seen in Figure 2, the dE/dx is low over much of its path. However, as it reaches the end of its range, the dE/dx more than doubles and correspondingly the RBE increases in this small area. This is where the low energy ionization events occur in Figure 5 and is the location of the greatest concentration of dose. Remember though, that because of the multitude of scattering events, the location of the high RBE portion occurs at random depths for each electron resulting in the green curve in Figure 2.

The culmination of all of these processes is, as shown in Figure 5, electrons causing ionizations and therefore depositing dose in the material.

## 4. Materials and Methods

In order to further support to these qualitative descriptions, computer simulations were performed characterizing the proportion of the various processes and to allow comparison of the energy spectra of the resulting electrons. The initialization involves defining the geometry of a tote system relative to each source of radiation. In the discussion below, the term "source assembly (SA)" will be used to refer to the elements of beam production before the radiation strikes the irradiated material. These are further defined below. In many large-scale irradiation facilities, the material to be irradiated is placed into "totes" of a uniform size to be transported through the irradiation area. Using realistic dimensions, the geometries of the radiation SAs and totes were defined for the three types of radiation to be as similar as possible so that the only appreciable difference would be the radiation type. Two simulations were run for each of the three types of radiation. One run had the SA and the totes; the other run had the SA only. The number of source particles (photons or electrons) simulated and the time for each run are listed in Table 1. Simulation summary data from the SA-only runs were subtracted from the SA-and-tote runs to calculate summaries of interactions that occurred in the totes alone. Some of the totes defined in the simulation were divided into slabs with the front surface of each slab used as a tally surface for accruing the energy spectra of the particles passing through that surface. This allowed tracking of the energy spectra of the photons and electrons as the particles progressed through the totes. This was done to see if there were any significant changes to the energies of the particles as a function of the depth of penetration. The evolution of the spectra of the photons and electrons through these slabs is the focus of this investigation.

|  | Electrons | | X-ray | | Gamma - Cobalt | |
| --- | --- | --- | --- | --- | --- | --- |
|  | SA Only | With Totes | SA Only | With totes | SA Only | With Totes |
| Particles | $1 \times 10^8$ | $6 \times 10^7$ | $5.4 \times 10^7$ | $7 \times 10^7$ | $1 \times 10^8$ | $1 \times 10^8$ |
| Minutes | 961 | 9890 | 15543 | 20445 | 1592 | 2827 |
| Days | 0.7 | 6.9 | 10.8 | 14.2 | 1.1 | 2.0 |

*Table 1 Number of particles simulated and elapsed time for Monte Carlo runs.*

For this work, the Monte Carlo program, MCNP6.2® was used. An MCNP® simulation provides information on all the processes that occurred for the photons and electrons and generated plots of the resulting energy spectra. These spectra could be tracked as the radiation penetrated through successive slabs of the irradiated material. However, in accounting for the various processes that result in these spectra, the program simply provides totals for the various processes throughout the simulation, so runs were performed with and without the irradiated material. By subtracting the two, the number of processes that occurred in the SA could be differentiated from those that occurred in the irradiated material.

The geometries that were constructed for each distinct irradiation technology (cobalt, x-ray, and electron) are shown in Figure 6. A tote system was assumed to represent the material to be sterilized. The totes were 122 x 122 x 61 cm (Width, Height, Depth) in size. They had 3 mm wall aluminum sides. The interior of each was a uniform mixture of air and polyethylene which had a density of 0.2 g/cm$^3$. This is a common average density for medical devices. Selected totes were divided into 12 slabs. The front surface of each of these slabs were defined as tally surfaces to provide information on the evolution of the energy spectra of the photons and electrons through the totes. In some totes, #6 in the gamma simulations and #2 in the electron and x-ray simulations, the slabs were perpendicular to the other totes as most of the propagation of the particles was due to scattering from totes that were more directly irradiated from the SA. This better followed the evolution of the energy spectra.

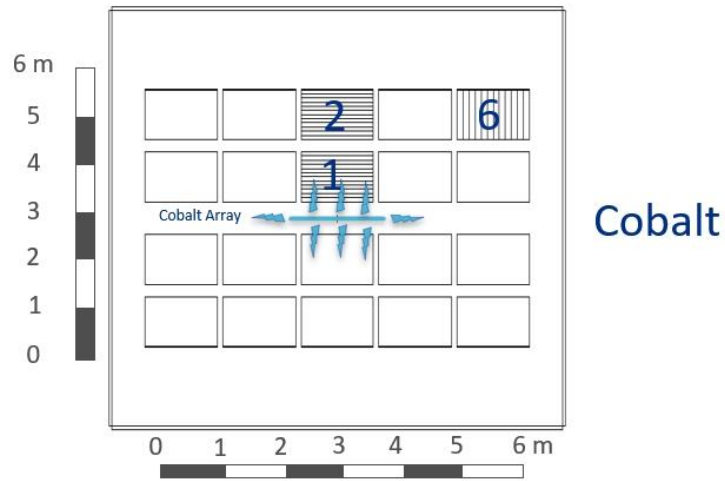
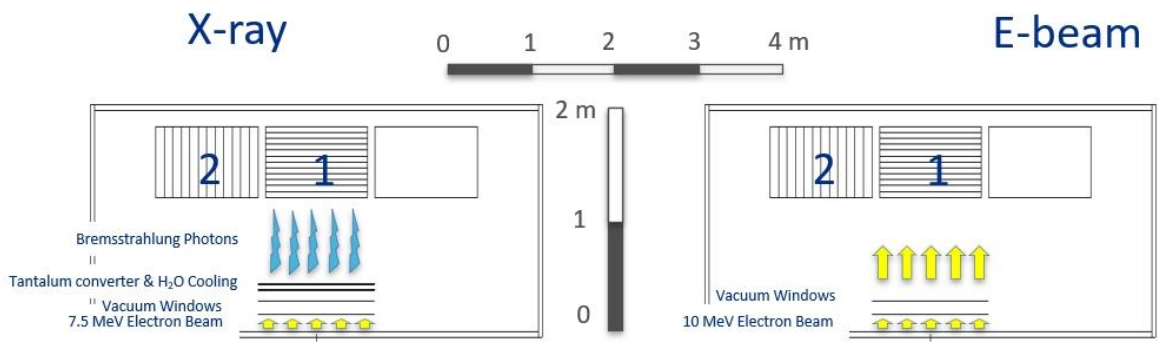

*Figure 6 Diagrams of the irradiation vaults for each radiation source that were simulated by the Monte Carlo program. Numbered totes are divided into 12 slabs to monitor the evolution of the energy spectra as the photons and electrons penetrate to deeper depths.*

Figure 7 provides more detail for the geometry of the SA. A tally surface was placed as noted in the figure. This counted the forward traveling photons and electrons emitted by the SA that were directed towards the objects to be irradiated.

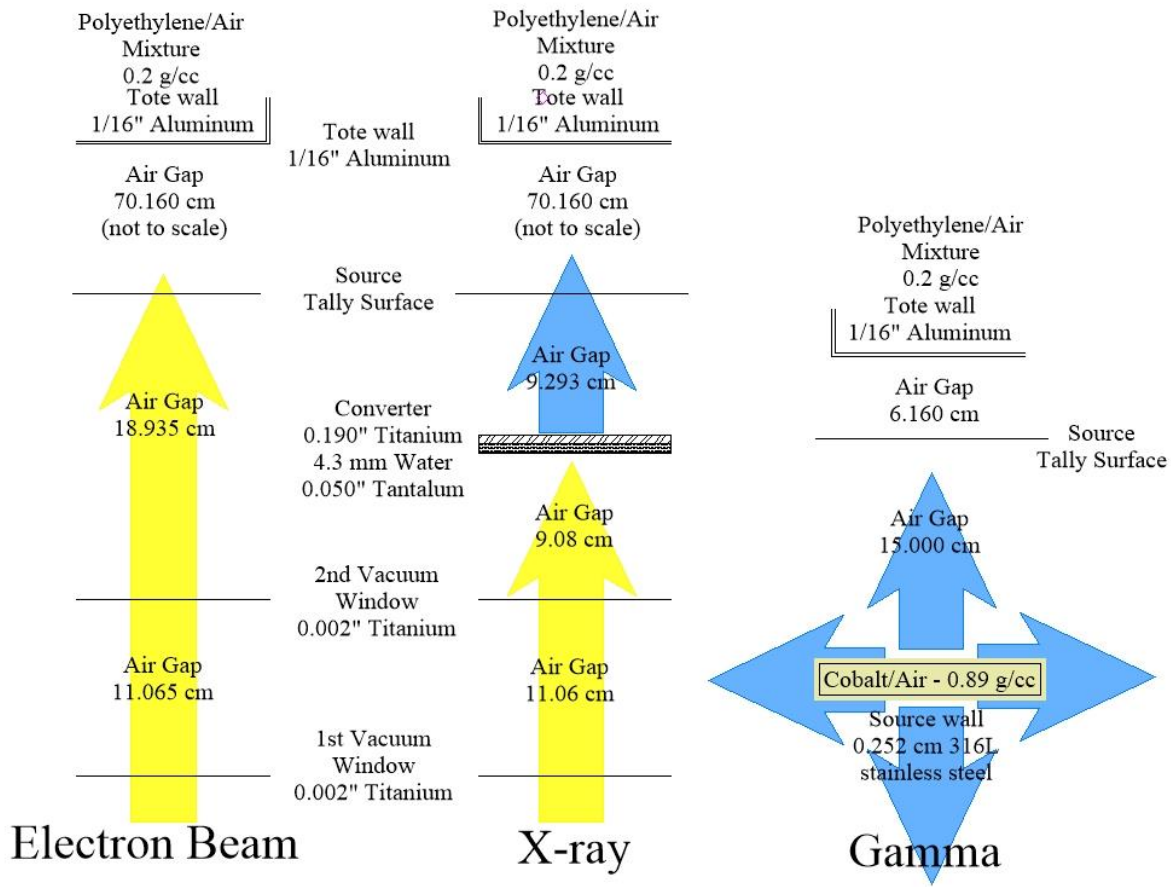

*Figure 7 Detail of geometry of electron and photon propagation from their source to the irradiated product. Vertical distances are to scale except where noted.*

For the gamma simulation, the SA included a homogeneous mixture of air and cobalt, to avoid the complexity associated with defining a large array of individual cobalt-containing pencils. This mixture was then placed in a rectangular box of stainless steel with dimensions of 120 cm (W) x 120 cm (H) x 60 cm (D) with wall thickness of 0.25 cm. The composition of the cobalt mixture and the stainless steel are listed in Table 2 along with the titanium used for the vacuum windows in the x-ray and electron cases. The mixture of materials in the cobalt SA uses the actual proportion of air and cobalt that are present in an actual array. The stainless-steel box represents the cladding of the cobalt elements and has a similar effect on the spectra as the gamma rays leave. The cobalt-air mixture and the stainless steel define the SA for the analysis below. Since the tally surface that measures the particles leaving the SA is set away, it is assumed that the combination of cobalt, air, and stainless steel provides an accurate spectrum of source energies. The source term of the Monte Carlo for the cobalt gammas was evenly distributed within the cobalt-air mixture and produced a 1.1732 MeV photon 49.97% and a 1.3325 MeV photon 50.03% of the time. The photons were emitted isotropically from all points within the SA volume, using the MCNP® defaults for a volume source.

| Cobalt Source – 0.89 g/cm³ | | Stainless Steel – 8 g/cm³ | | Titanium – 4.43 g/cm³ | |
|---|---|---|---|---|---|
| Isotope | Wt % | Isotope | Wt % | Isotope | Wt % |
| N-14 | 827 ppm | C-12 | 200 ppm | H-1 | 100 ppm |
| O-16 | 254 ppm | N-14 | 0.1 | C-12 | 500 ppm |
| Co-59 | 88.9 | Si-28 | 0.75 | N-14 | 200 ppm |
| Co-60 | 10.9 | P-31 | 450 | O-16 | 200 ppm |
| | | S-32 | 300 | Ti* | 98.6 |
| | | Cr-52 | 17 | Fe* | 0.2 |
| | | Mn-55 | 2 | Ni* | 0.75 |
| | | Fe* | 65.6 | Mo* | 0.3 |
| | | Ni* | 12 | | |
| | | Mo* | 2.5 | | |
| Cobalt/air mixture | | * uses naturally occurring abundances | | | |

*Table 2 Elemental composition of materials used source configuration for computer simulations with MCNP®.*

The SA for the electron beam consists of two titanium windows (composition noted in Table 2) which were 50 µm thick, spaced 11 cm apart. The x-ray SA adds a pure tantalum converter and a water-cooling channel and another titanium window. The tantalum is 0.127 cm thick and is 20 cm downstream of the second titanium window. The water channel abuts the tantalum and is 0.43 cm thick. The final titanium window to contain the water is 0.843 cm thick.

The electrons for both electron and x-ray simulations were generated from a plane, 126 cm wide by 1 m tall. The direction was perpendicular to the first titanium window. The width spans from the midpoint of the gap between totes 1 and 2 to the midpoint of the gap between tote 1 and the unnumbered tote. This was done to simulate the movement of the irradiated material in front of a thin beam of radiation. This simulates the accumulated dose of a scan along the path of movement. The initial energy of the electrons for the electron simulation was a central energy of 10 MeV and a sigma of 0.018 MeV using the MCNP® source probability function of -4 for a gaussian distribution of energy. For the x-ray simulations, the central energy was 7.5 MeV, also with a sigma of 0.018 MeV. Again, a tally surface was placed between the source components and the totes to measure the incident electrons and x-rays moving in the forward direction.

The simulations were run on a single core of a 2.11 GHz Intel(R) Core(TM) i7-8650U CPU. The target number of source particles was $1\times10^8$ however some runs fell short of this due to unexpected interruptions. Table 1 lists the actual number of particles for each run and the elapsed time. One can see the significantly increased run time due to simulating the bremsstrahlung process for the generation of the x-rays. No variance reduction methods were used. Surface tally F1 was used for the source tallies. Tally F1 was also used for the tallies in the totes using the front surface of each slab as the tally surface (or the central-axis facing surface for the perpendicular slabs). The physics mode simulated only photons and electrons using the default MCNP® values for both. For photons this included thick-target bremsstrahlung photons, coherent scattering, and Doppler energy broadening. It excluded photonuclear and photofission phenomenon. For electrons it included thick-target analog bremsstrahlung, photon production, full bremsstrahlung tabular angular distribution, sampled value electron straggling, knock-on electron creation, Goudsmit-Sunderson angular deflection for Coulomb scattering, and a 1 keV threshold for single-event transport.

# 5. Results

## 5.1 Particle Quantity Comparisons

Figure 8a repeats Figure 1 for comparison with Figure 8b. Figure 8b shows the energy spectra of both the photons (solid lines) and the electrons (dashed lines) that were generated upon leaving the source structures and passing through the first tally surface. This more accurately describes the initial radiation environment that irradiated products are exposed to.

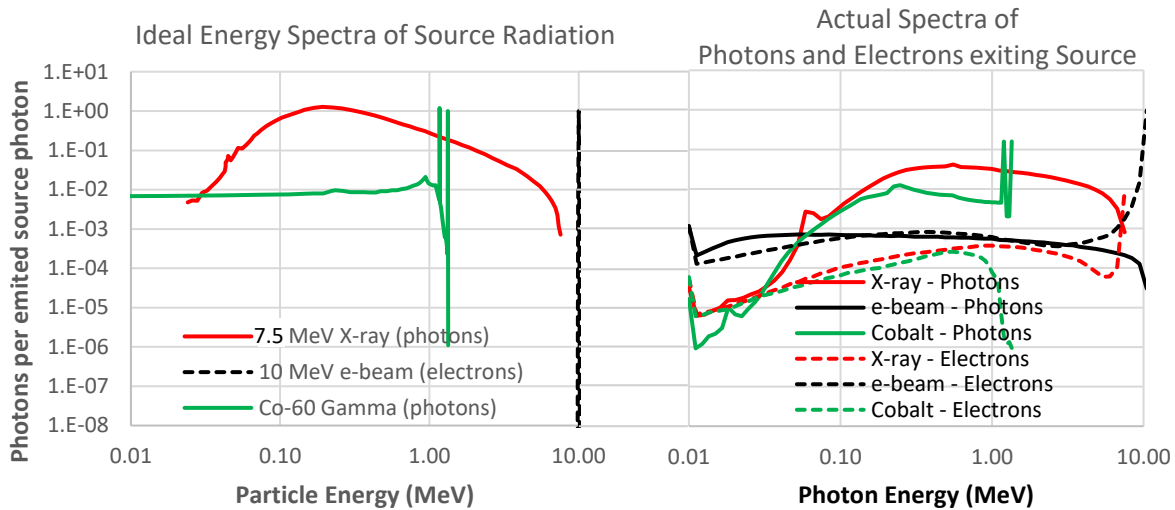

*Figure 8 a) Left – ideal spectra of source particles. X-ray spectra generated using parameters from reference [13]. b) Right – Actual spectra as particles leave the source. This includes all processes within the cobalt array, and the vacuum and bremsstrahlung converter. The actual spectra show that no radiation source provides single identity particles. Electrons are dashed; photons are solid lines.*

Even before the radiation strikes the irradiated objects, it consists of a mixture of electrons and photons. In the cobalt source, there is scattering in the air and the steel cladding. In the x-ray simulation, the electron beam from the accelerator strikes the tantalum converter in which x-rays are produced via bremsstrahlung. In addition, there are vacuum windows and water cooling that the electrons and x-rays must pass through before reaching the totes. In the electron case, the very monochromatic electrons, must pass through two thin metal vacuum windows.

The point of this figure is to illustrate that, once the beam has emerged from all of the materials that comprise the source, in no case is there a pure beam of one type of particle nor are they monoenergetic.

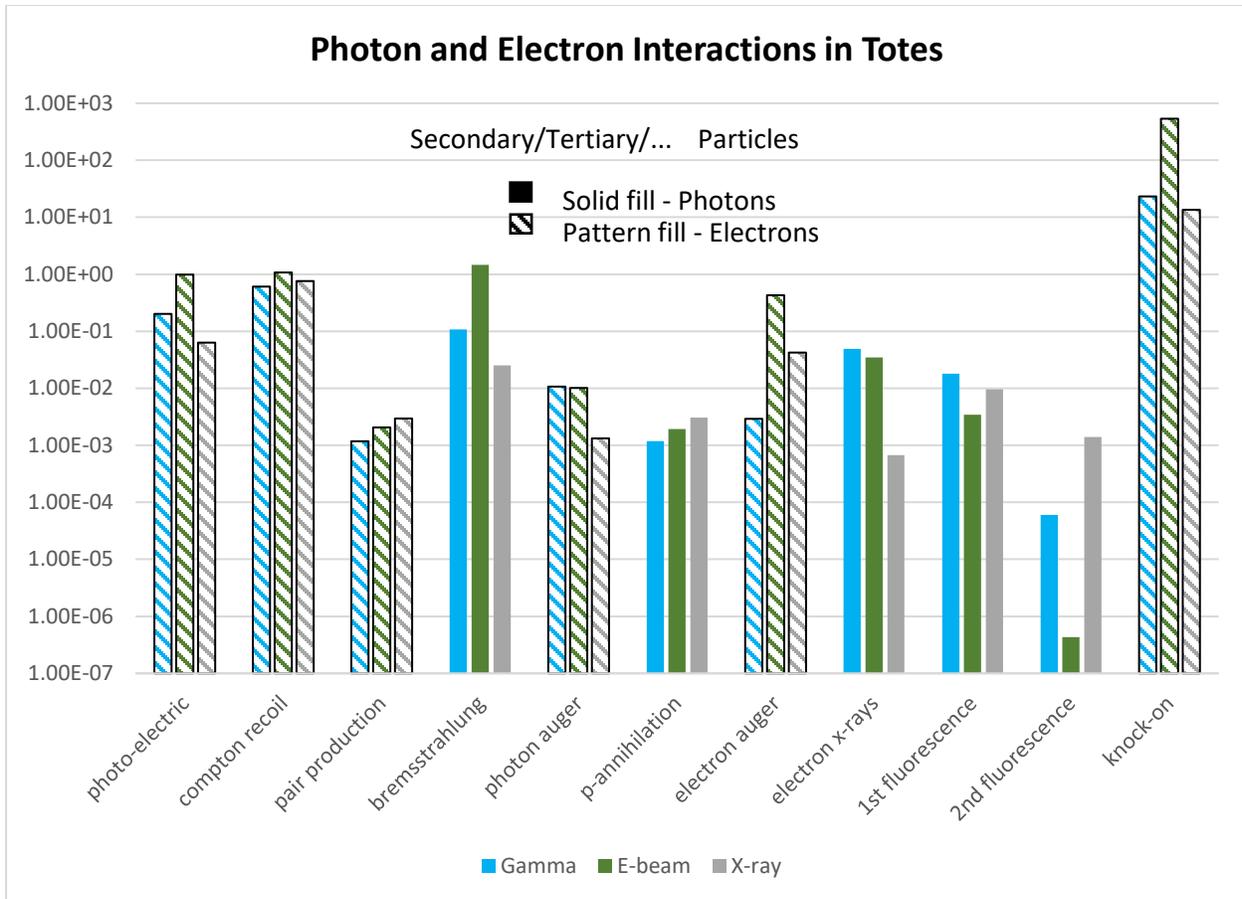

*Figure 9 Normalized quantities of photons and electrons created within the irradiated totes as incident radiation penetrates the totes. Normalized to one emitted photon or electron (filled circles on left). Illiustrates the relative number of photons or electrons created by each of the possible interaction processes. Electron and photon quantities to the left of the dashed line are additional particles created within the source by the initial source particles.*

The MCNP® output provides a summary of the processes that create and eliminate photons and electrons that occur during the simulation. These summaries are distinct from the tallies that were used to generate Figures 10 - 12 later in this report. The summaries report the number of occurrences of each process for the entire simulation volume. Each of the three radiation technologies was run twice, once with the SAs and the totes in the simulation volume and the other run without the totes but the same SAs and simulation volume. As was seen in Table 1, the number of source particles for a run ranged between $5.4 \times 10^7$ and $1 \times 10^8$. The summaries for each run were normalized to $1 \times 10^8$. The normalized results of runs without totes were subtracted from those with totes and then divided by $1 \times 10^8$, giving the relative activity that occurred in the totes normalized to a single emitted photon or electron. Figure 9 summarizes all of these processes occurring within the totes. These data are also listed in Table 3.

Moving from left to right in Figure 9 progresses from interactions involving predominately high energy particles to interactions involving low energy particles. To the left we see the three primary initial interactions that produce electrons, the photo-electric effect, Compton scattering, and pair production. Of these, Compton scattering is the most prevalent however, in the electron source case, the photo-electric effect is comparable. Skipping to the middle of the figure we see the annihilation of the positrons that were produced in pair production.

|                      | Gamma        | Electron     | X-ray        |                        |
|----------------------|--------------|--------------|--------------|------------------------|
| Source – Photons     | 1.00         | -            | -            |                        |
| Source - Electrons   | -            | 1.00         | 1.00         |                        |
| Photoelectric        | 0.202        | 0.986        | $6.33 \times 10^{-2}$ |               |
| Compton Scattering   | 0.608        | 1.07         | 0.760        |                        |
| Pair Production      | $1.18 \times 10^{-3}$ | $2.06 \times 10^{-3}$ | $2.95 \times 10^{-3}$ |       |
| Bremsstrahlung       | 0.107        | 1.46         | $2.51 \times 10^{-2}$ |               |
| Photon Auger         | $1.07 \times 10^{-2}$ | $1.01 \times 10^{-2}$ | $1.32 \times 10^{-3}$ | Particles generated in the Totes |
| Positron Annihilation| $1.18 \times 10^{-3}$ | $1.94 \times 10^{-3}$ | $3.05 \times 10^{-3}$ |       |
| Electron Auger       | $2.91 \times 10^{-3}$ | 0.429        | $4.23 \times 10^{-2}$ |       |
| Electron X-rays      | $4.89 \times 10^{-2}$ | $3.46 \times 10^{-2}$ | $6.76 \times 10^{-4}$ |       |
| 1st Fluorescence     | $1.81 \times 10^{-2}$ | $3.46 \times 10^{-3}$ | $9.61 \times 10^{-3}$ |       |
| 2nd Fluorescence     | $5.98 \times 10^{-5}$ | $4.31 \times 10^{-7}$ | $1.40 \times 10^{-3}$ |       |
| Knock-on             | 23.2         | 535          | 13.5         |                        |

*Table 3 Distribution of electron and photon creation processes by irradiation technology. Yellow denotes photon creating processes, blue denotes electron creating processes. All values are normalized to one source photon or electron.*

To the right (or below in Table 3) of the first three processes (photoelectric, Compton scattering, pair production) are other processes that produce more photons and electrons. Next comes bremsstrahlung where many of the electrons produce more photons as do electron x-ray and fluorescence processes. Photons and electrons can liberate more electrons though the Auger process. The Auger process is another high dE/dx, high RBE process that results in high dose density. Some interactions excite atoms and as their electrons settle back into the ground state, they release x-rays. Fluorescence creates more photons. Knock-on electrons (delta-rays) are electrons created by ionization from electron impacts.

In the end, there are at least an order of magnitude greater number of electrons than originally sourced. This represents the processes that perform the biological damage that leads to sterility. Again, note that values are normalized to one initial source photon or electron and are not meant to compare dose between the three technologies.

Now we can see the complexity and interplay of all the processes described. While the photoelectric, Compton, and pair production electrons distribute themselves throughout the material, it is the Auger and knock-on electrons at the terminal end of the cascade that provide the bulk of the ionization and dose. These are the blobs, short tracks, and terminating blobs in Figure 5. The bremsstrahlung x-rays, electron x-rays, and annihilation and fluorescence photons generated by medium energy electrons also serve to distribute energy throughout the medium.

The complexity of events and the energy of the processes that produce sterilization relative to the energy of the source particles are what decouples sterilization from the type of, and initial energy of, the original radiation.

### 5.2. Results – Particle Spectra Comparisons

Finally, we look at the energy spectra of the photons and electrons in the totes for all three radiation sources (Figures 10, 11 and 12). Figure 10 shows the spectra of photons and electrons for all three source types over the full range of energy. Figure 11 looks at the photons for each source and focuses on the energy range below 500 keV. Figure 12 looks at the electrons below 500 keV. Assuming that biologically effective work is done by electrons with energy of 500 keV (the average energy of a Compton Electron) and below, that area has been identified as the "kill zone" on the plots.

Not surprisingly, there are noticeable differences in the photon spectra, given their spectra as they exited their source (Figure 10, a-c). As noted previously, some totes were divided into 12, 2" thick, slabs in order to follow the evolution of the energy spectra as the radiation penetrates deeper into the totes and from tote to tote. The detailed photon plots of Figure 11 show that there is some evolution of the photon spectra as the photons penetrate into the totes. The peak of the distribution moves to lower energies at deeper depths. The distribution in totes 1 and 2 are shown for the cobalt case (Figure 11 a) since they are two totes deep as in many actual irradiation chambers. One can see that the attenuation continues with just a small increase in attenuation due to the tote walls and air gap between the totes. One can also see in Figures 10 a-c that the totes that are off to the side see mostly scattered radiation and the number of photons that these totes are exposed to at those locations are at least an order of magnitude less. For this reason, Figures 11 and 12 do not include the side totes.

Things are much more uniform in the electron plots, particularly in the x-ray and cobalt cases (Figures 10 d-e). Figure 12 shows more detail at electron energies lower than 500 keV.

Again, we see the steady attenuation with depth of the spectra in the cobalt case (Figure 12 a). For the x-ray case, there is a constancy in the first three layers as the low energy photons that are generating the electrons are preferentially attenuated at these shallower depths (Figure 12 b). In the case of the electron beam (Figure 12 c), one can see that the penetration of the electron beam is only about 10 cm (5 slabs). There is then a drop-off of a factor of 10 in passing to the next slab. The electrons that do exist in deeper slabs and in the following tote are due to electrons generated by photons that were either generated in the vacuum windows of the source or generated within the totes through the photon production processes mentioned above.

The advantage that x-rays have over gamma rays from cobalt-60 in providing a better dose uniformity ratio (DUR) can be seen in Figures 11 and 12. There is less attenuation between slabs 1 and 12 in tote 1 in both the photon and electron spectra.

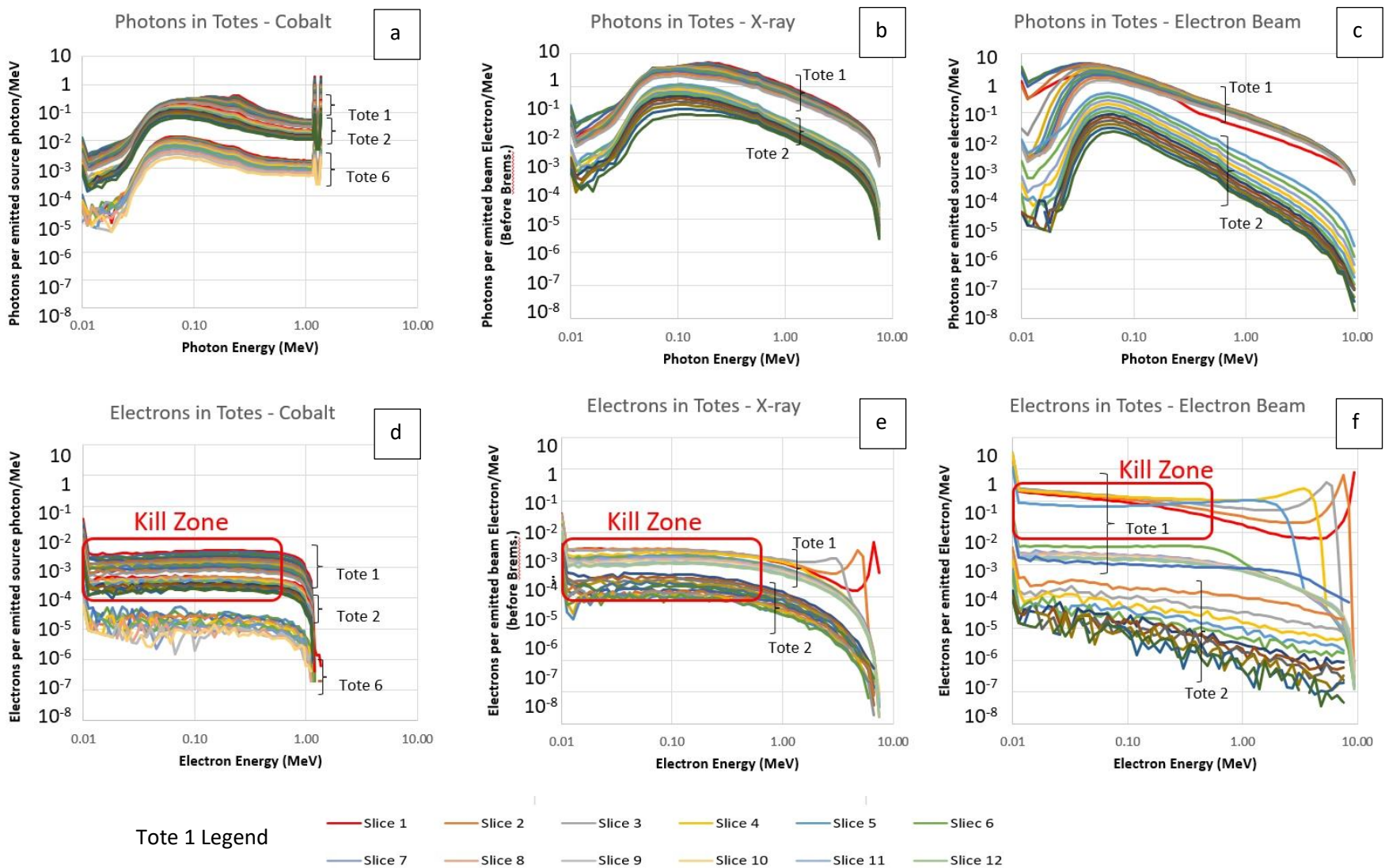

*Figure 10 Energy spectra of photons (top row) and electrons (bottom row) in the segmented totes simulated by MCNP®. Each of the totes was divided into 12 segments in order to see the evolution of the energy spectrum as the particles penetrated deeper into the tote system. Tote numbers refer to Figure 6. Color coding of slices for Tote 1 shown in legend below the figure.*

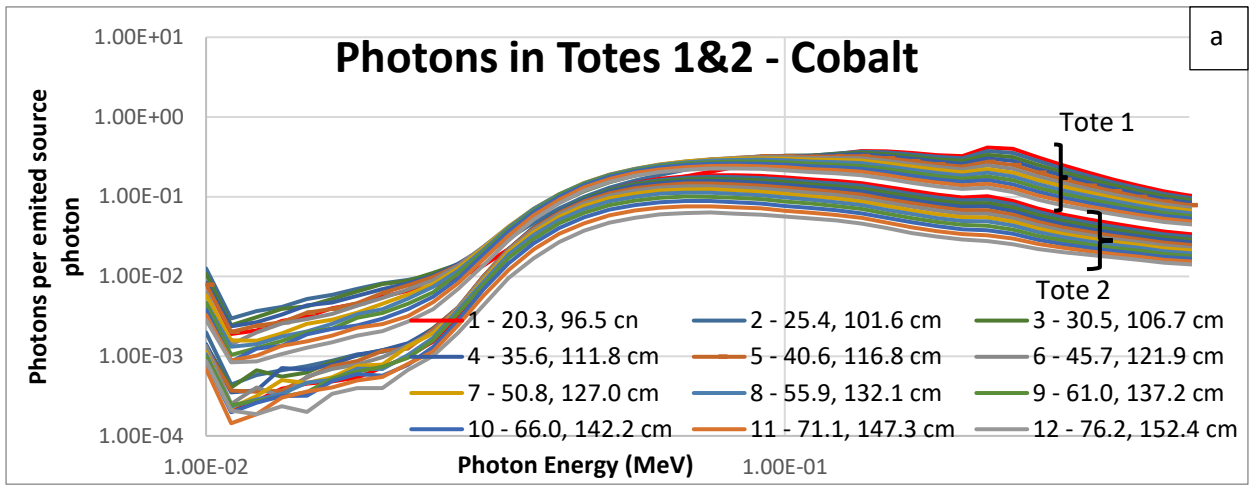
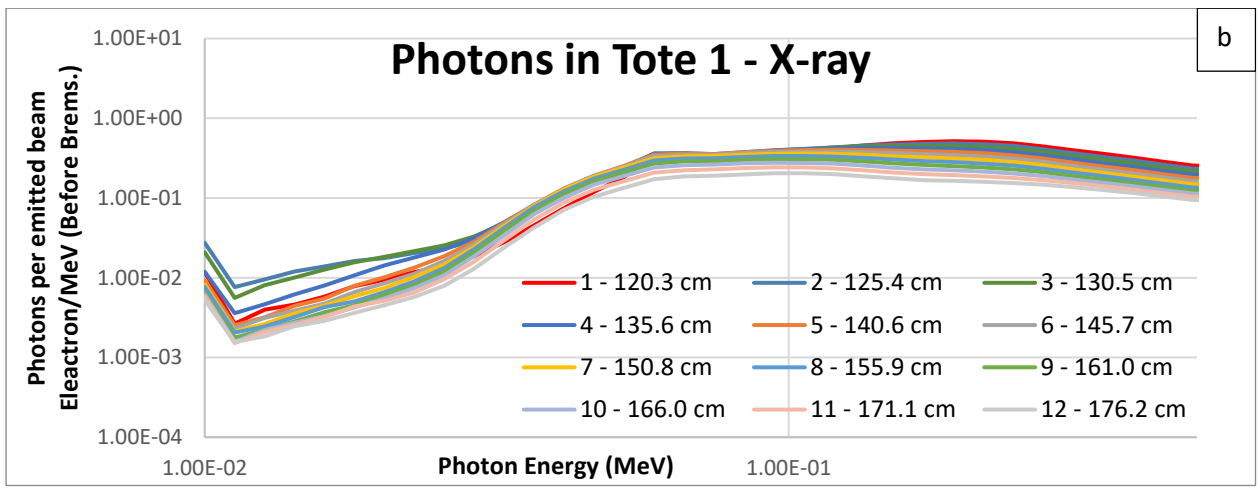
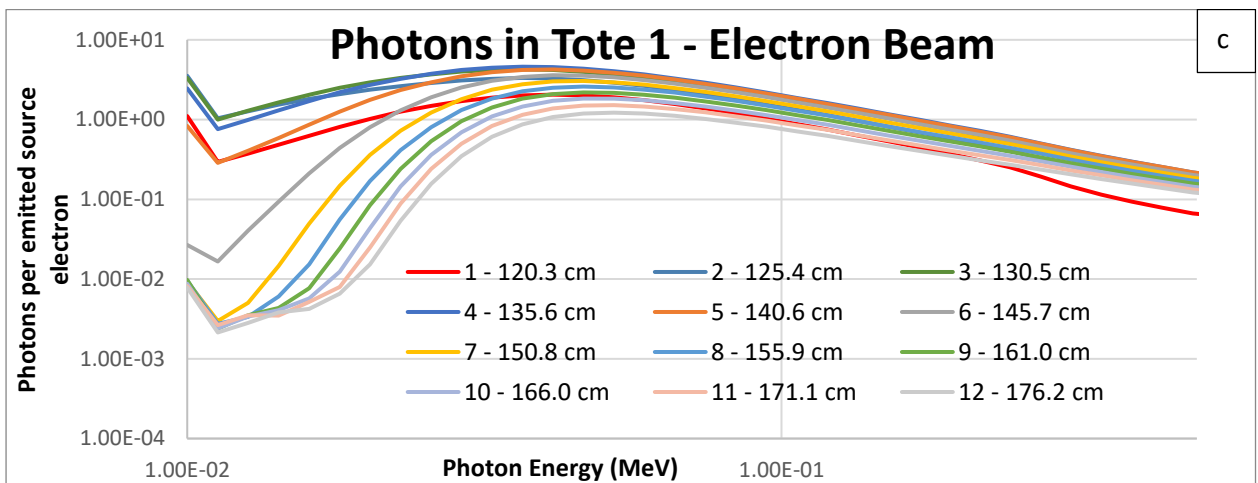

Figure 11 Detail of the evolution of photon energy, between 10 and 500 keV, in Tote 1 for each of the three radiation sources.

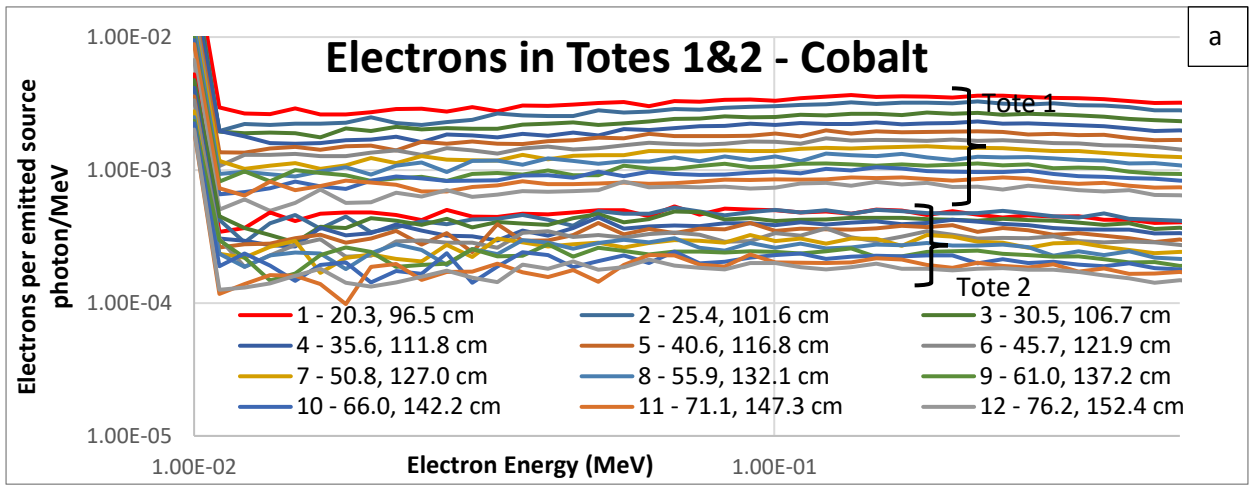
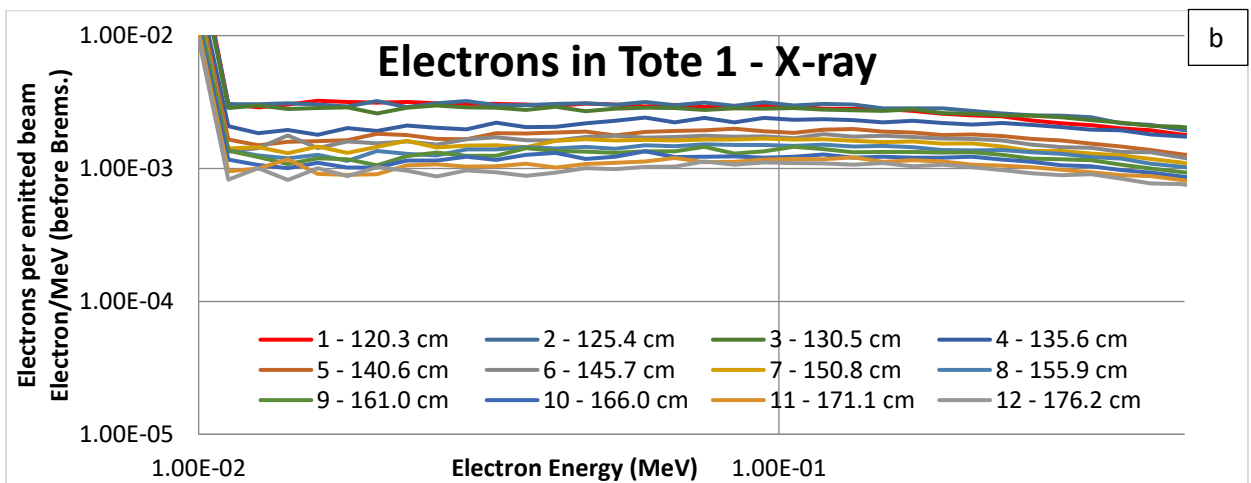
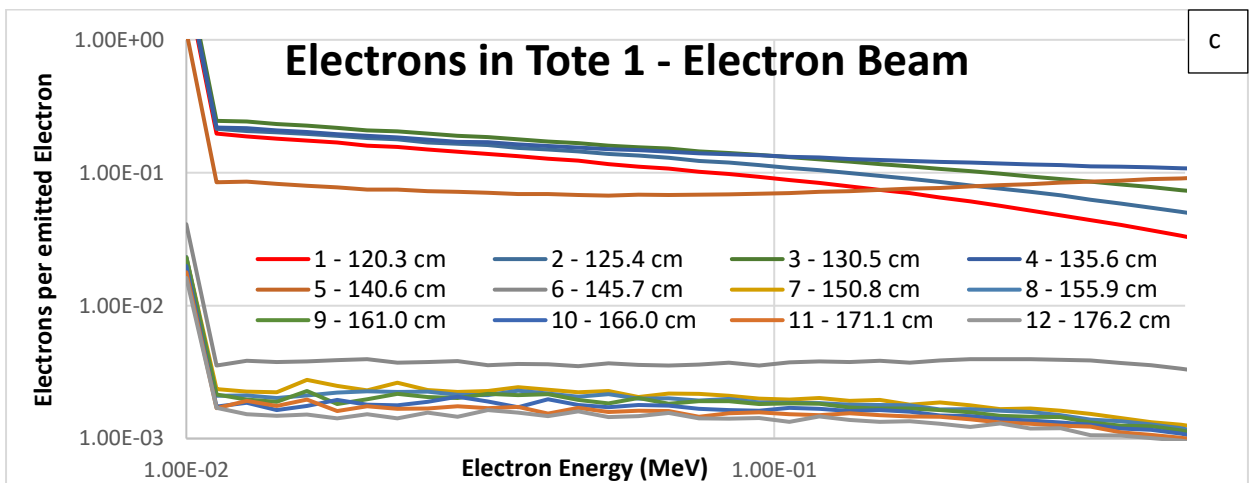

Figure 12 Detail of the evolution of electron energy, between 10 and 500 keV, in Tote 1 for each of the three radiation sources.

It is seen above that there is very little difference in the electron energy spectra between the three irradiation technologies. This study did not investigate the effect of inhomogeneities, that is, variations in density of actual products rather than the uniform density assumed here. Such variations can cause variations in dose to specific locations. It could change the balance of photons vs electrons. But given the flat energy response of the results seen above, there is nothing to suggest that any of the three radiation sources would be affected differently than the others. Looking at the theories for the cross sections (which determine the probabilities of a particular interaction) of the three initial processes of interaction, the photoelectric effect, Compton scattering, and pair production, no dependence is seen that could be attributed to the radiation source. Any material dependencies in these processes are properties of the irradiated material, not the incident radiation. Therefore, the spectra of the incident radiation have no bearing on the processes that occur.

As mentioned above the energy of the initial radiation in the incident beam is orders of magnitude greater than the energies of the ionization events that constitute the delivery of dose. The image that this work is trying to convey is that high-energy particles, predominately photons, transport energy into material and in depth. As this is happening, a cascade is forming of photons and electrons which grows in particle density as the mean energy of the particles is degraded. Ultimately, low energy electrons are generated which locally deposit energy and generate dose. The simulations show that the spectra of energies in each of the three irradiation technologies, and particularly in comparing x-ray to gammas from cobalt-60, show no difference. This is especially true in the energy range, below 500 keV, where the dose accumulation occurs.

When looking at standards and guidance documents such as ISO 11137 and AAMI TIR 104, there are many factors to consider in understanding the performance of a sterilization procedure for a particular radiation technology. In addition to absolute value of dose, which has been shown here to be indistinguishable, dose uniformity and dose rate must also be considered. As seen in Figure 3, the penetration of electrons is significantly less since they are starting out as charged particles. Gamma and x-rays are both photons and therefore have greater penetrating ability. X-rays used for sterilization are typically between 5 and 7.5 MeV. While x-rays have a wide spectrum of energies, the large fraction above the energy of gamma rays from cobalt-60 gives x-rays greater penetrating power and can be exploited to give better dose uniformity than gamma rays.

## 6. Conclusion

When one looks at the spectrum of photons from the decay of Cobalt-60, the sharpness of the two characteristic energies, 1.17 and 1.33 MeV, gives the impression of crispness and precision with regard to their impact on delivering dose to items being sterilized. In contrast, the broad spectrum of the bremsstrahlung photons produced from an electron beam onto a converter with high atomic number can give the impression of disorder and complexity in understanding the delivered dose.

On the basis of the above data and discussions, it is concluded that there is no difference in the electron energy spectra and dose deposition between the three radiation sources used for industrial sterilization: gamma rays from the decay of Cobalt-60, x-rays produced from an electron beam striking a bremsstrahlung converter, and direct electron beams. Despite apparent differences in the initial spectra and the fact that the initial radiation is photon-based in two cases and electron-based in the third, the magnitude of the difference between the energy of the source particles and the energy of the particles that deliver dose provides the opportunity to form an ever-growing cascade of both photons and electrons resulting in moderate to low energy electrons which deliver their energy locally. The higher energy photons and electrons penetrate the irradiated material before the low energy electrons deliver the dose. The spectra of the electrons below 500 keV from all three sources is uniform from 10 keV to 500 keV. The results from the electron beam simulation do exhibit the limited range compared to the two photon sources. In comparing the spectra from x-rays with the spectra from cobalt gammas the advantage in DUR of the former can be seen. But if one were to be presented with the spectrum from an individual slab, it would be difficult to identify if the source was x-ray or gamma ray.

This work only examined the composition and energy spectra of the particles that deliver dose. The three radiation technologies have dose rates that can vary by a factor of 1,000. Material effects due to dose rate may occur due to induced temperature profiles or the atmosphere surrounding the irradiated material. The reader is referred to the continuing work of Team Nablo [7-10] to follow up on these aspects.

It is hoped that this work will assist in the understanding of the equivalency of these three radiation sources for radiation sterilization.

# 7. Acknowledgement

The author gratefully acknowledges the helpful comments and review of this work by James Hathcock and Byron Lambert.

Work supported by the Fermi National Accelerator Laboratory, managed and operated by Fermi Research Alliance, LLC under Contract No. DE-AC02-07CH11359 with the U.S. Department of Energy. The U.S. Government retains and the publisher, by accepting the article for publication, acknowledges that the U.S. Government retains a non-exclusive, paid-up, irrevocable, world-wide license to publish or reproduce the published form of this manuscript, or allow others to do so, for U.S. Government purposes.